\documentclass[12pt,oneside,reqno]{amsart}	
\usepackage[margin=1in]{geometry}			
\usepackage[parfill]{parskip}					
\usepackage{graphicx}
\usepackage{amssymb}
\usepackage{epstopdf}
\usepackage{cite}			
\usepackage[hidelinks]{hyperref} 
\title{Causal intuition and delayed-choice experiments}
\author{Michael B. Heaney\\3182 Stelling Drive\\Palo Alto, CA 94303\\mheaney@alum.mit.edu}
\date{26 December 2020}						
\begin{document}
\maketitle
\begin{abstract}
The conventional explanation of delayed-choice experiments appears to violate our causal intuition {at the quantum level}. I reanalyze these experiments using time-reversed and time-symmetric formulations of quantum mechanics. {The time-reversed formulation does not give the same experimental predictions. The time-symmetric formulation gives the same experimental predictions but actually violates our causal intuition at the quantum level.} I explore the reasons why our causal intuition may be wrong at the quantum level, suggest how conventional causation might be recovered in the classical limit, propose a quantum analog to the classical block universe viewpoint, and speculate on implications of the time-symmetric formulation for cosmological boundary conditions.
\end{abstract}
\section{Introduction}\label{sec1}
One of the grand challenges of modern physics is to resolve the conceptual paradoxes in the foundations of quantum mechanics~\cite{Smolin}. Some of these paradoxes concern our causal intuition. For example, in 1926 Lewis proposed a delayed-choice thought experiment which appeared to show retrocausation in the Conventional Formulation of quantum mechanics~\cite{LewisA, LewisB}. Retrocausation, also known as future input dependence~\cite{Wharton}, is when a model parameter associated with time $t$ depends on model inputs associated with times greater than $t$. He considered a double-slit interference experiment using a single photon from a distant star. One thousand years after the photon has left the star, but just before it reaches the two slits ($A$ and $B$) on Earth, we randomly choose to either keep both slits $A$ and $B$ open, or intervene to close slit $A$ only, or intervene to close slit $B$ only. We repeat this experiment for a large number of single photons to obtain an ensemble of experimental results. In the sub-ensemble where we chose to keep both slits open, we see an interference pattern, implying each photon took both routes from the star. In the sub-ensemble where we intervened to keep only one slit open, we do not see an interference pattern, implying each photon took only one route from the star. Lewis concluded that "in some manner the atom in the source $S$ can foretell before it emits its quantum of light whether one or both of the slits $A$ and $B$ are going to be open". Our intervention appears to cause the photon to change its route before the intervention actually happens, in violation of our causal intuition that effects never happen before interventions. This is the delayed-choice paradox. Weizs\"{a}cker and Wheeler later rediscovered and elaborated on Lewis's thought experiment~\cite{WeizsackerA, WeizsackerB, WheelerA, WheelerB}. This apparent paradox has been confirmed in experiments with photons, neutrons, and atoms~\cite{HellmuthA,HellmuthB,BaldzuhnA,LawsonDakuA,KawaiA,KimA,JacquesA,PeruzzoA,TangA,RoyA,KaiserA}. The most recent review of delayed-choice experiments says "It is a general feature of delayed-choice experiments that quantum effects can mimic an influence of future actions on {past events}"~\cite{Ma}. {The word "mimic" is used because the authors ascribe to an interpretation of the wavefunction as only a "catalog of our knowledge", not a real physical object. There is then no delayed-choice paradox. This is the conventional Copenhagen interpretation~\cite{Commins,CTDL,Griffiths}. I have instead chosen to interpret the wavefunction as a real physical object.} Note that there are many different formulations and interpretations of quantum mechanics, but this paper will only be concerned with the Conventional, Time-Reversed, and Time-Symmetric Formulations.

The structure of this paper is as follows. {Section \ref{sec2} describes how Wheeler analyzed the delayed-choice experiment using the Conventional Formulation (CF) of quantum mechanics. Section \ref{sec3} describes a reanalysis of the same experiment using a Time-Reversed Formulation (TRF) of quantum mechanics. Section \ref{sec4} describes a reanalysis of the same experiment using a Time-Symmetric Formulation (TSF) of quantum mechanics. Section \ref{sec5} discusses the results and draws conclusions.}
\section{The Conventional Formulation of the Delayed-Choice Experiment}\label{sec2}
Let us consider thought experiments with the neutron Mach--Zehnder interferometer (MZI) shown in \mbox{Figure~\ref{fig1}}, where a single neutron is emitted from either source $S1$ or source $S2$. The Conventional Formulation (CF) postulates that a single free particle with mass $m$ is described by a wavefunction $\psi(\vec{r},t)$ which satisfies the initial conditions and evolves in time according to the Schr\"{o}dinger equation:

\begin{equation}
i\hbar\frac{\partial\psi}{\partial t}=-\frac{\hbar^2}{2m}\nabla^2\psi.
\label{2}
\end{equation}
\begin{figure}[htbp]
\centering
\includegraphics[width=3.9in]{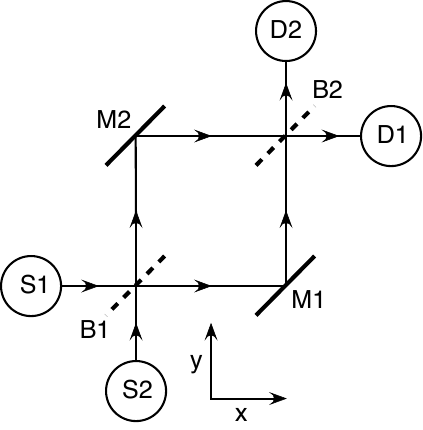}\label{fig1}
\caption{The neutron Mach--Zehnder interferometer (MZI). $S1$ and $S2$ are neutron sources, $B1$ and $B2$ are beam-splitters, $M1$ and $M2$ are mirrors, and $D1$ and $D2$ are detectors. The sources $S1$ and $S2$ can each emit a single neutron on command. The MZI is constructed such that neutrons emitted from $S1$ are always detected at $D1$, while neutrons emitted from $S2$ are always detected at $D2$.}
\label{fig1}
\end{figure}
We will use units where $\hbar=1$ and assume the wavefunction $\psi(\vec{r},t)$ is a traveling gaussian with an initial standard deviation $\sigma=50$, momentum $k_x=0.4$, and mass $m=1$. 

{\mbox{Figure~\ref{fig2}} shows how the neutron's CF probability density $\psi^\ast\psi$ evolves over time in the MZI, assuming the initial condition is localization in source $S1$. At $t=0$, $\psi^\ast\psi$ is localized inside the source $S1$. At $t=3000$, $\psi^\ast\psi$ has been split in half by beam-splitter $B1$, and the halves are traveling towards mirrors $M1$ and $M2$. At $t=5000$, the two halves have been reflected by $M1$ and $M2$ and are both traveling towards beam-splitter $B2$. At $t=7000$, the two halves have been recombined by $B2$, with $\psi^\ast\psi$ interfering constructively towards detector $D1$ and destructively towards detector $D2$. At $t=8000-\delta t$, $\psi^\ast\psi$ arrives at $D1$, but is not localized inside $D1$. Upon measurement at $t=8000$, $\psi$ collapses to a different wavefunction $\xi$, with $\xi^\ast\xi$ localized inside $D1$. Similarly, if $\psi^\ast\psi$ had been localized inside the source $S2$ at $t=0$, it would have taken both routes and collapsed to being localized inside $D2$ at $t=8000$.}
\begin{figure}[htbp]
\centering
\includegraphics[width=5in]{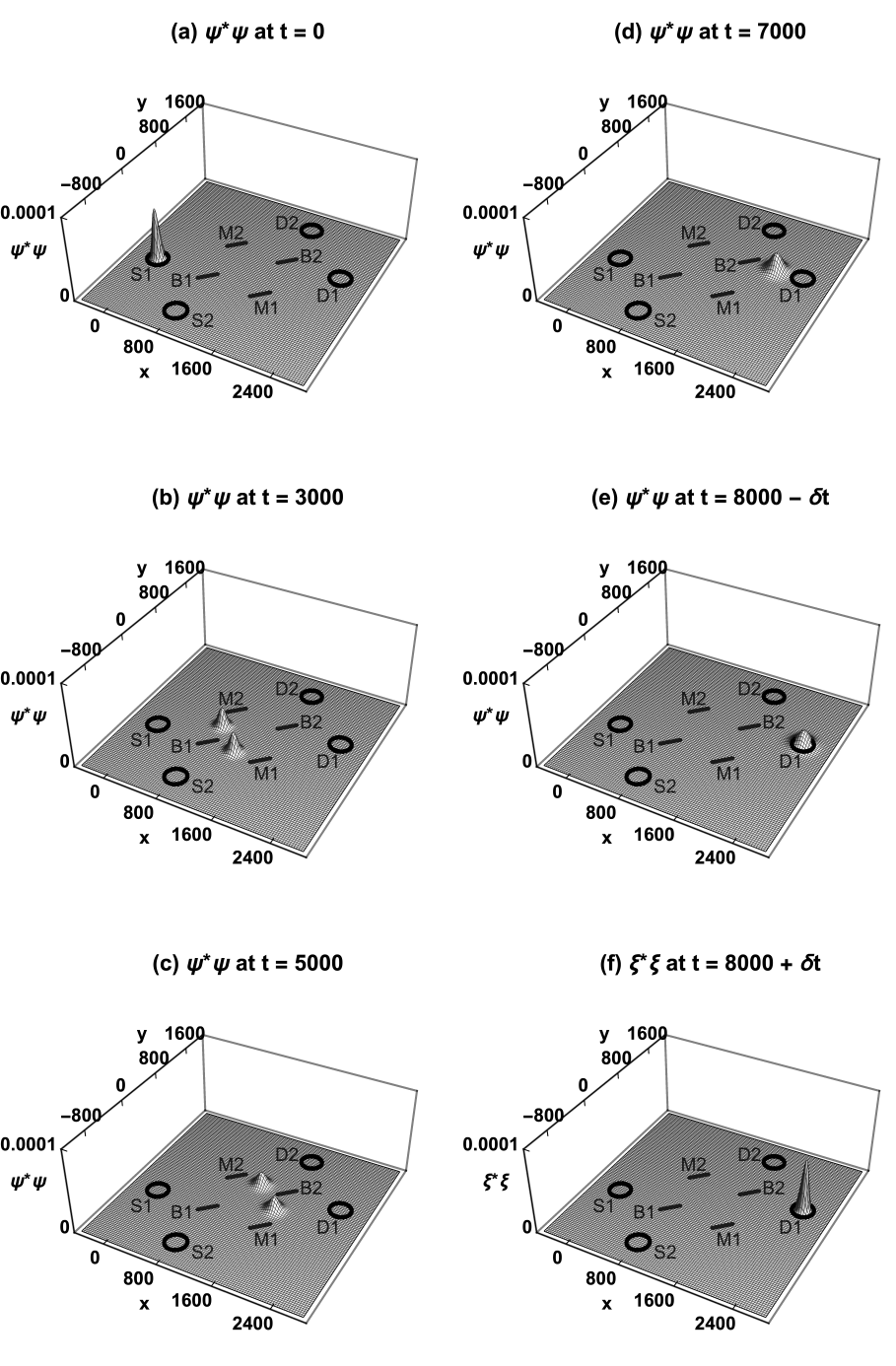}\label{fig2}
\caption{The Conventional Formulation (CF) of the MZI experiment, with a single neutron emitted from $S1$. (\textbf{a}) The probability density $\psi^\ast\psi$ is localized inside $S1$. (\textbf{b}) $\psi^\ast\psi$ has been split in half by $B1$. (\textbf{c})~The two halves have been reflected by $M1$ and $M2$. (\textbf{d}) The recombined $\psi^\ast\psi$ interferes constructively towards $D1$ and destructively towards $D2$. (\textbf{e}) $\psi^\ast\psi$ arrives at $D1$, but is not localized inside $D1$. (\textbf{f}) Upon measurement at $t=8000$, $\psi$ collapses to a different wavefunction $\xi$, localized inside $D1$. {Wavefunction collapse is a postulate of the CF, and introduces a time-asymmetry.}}\label{fig2}
\end{figure}
\mbox{Figure~\ref{fig2}} shows how the particle's CF probability density $\psi^\ast\psi$ evolves over time in the MZI, assuming the initial condition is localization in source $S1$. At $t=0$, $\vert\phi^\ast\psi\vert$ is localized inside the source $S1$. At $t=3000$, $\psi^\ast\psi$ has been split in half by beam-splitter $B1$, and the halves are traveling towards mirrors $M1$ and $M2$. At $t=5000$, the two halves have been reflected by $M1$ and $M2$ and are both traveling towards beam-splitter $B2$. At $t=7000$, the two halves have been recombined by $B2$, with $\psi^\ast\psi$ interfering constructively towards detector $D1$ and destructively towards detector $D2$. At $t=8000-\delta t$, $\psi^\ast\psi$ arrives at $D1$, but is not localized inside $D1$. Upon measurement at $t=8000$, $\psi$ collapses to a different wavefunction $\xi$, with $\xi^\ast\xi$ localized inside $D1$. Similarly, if $\psi^\ast\psi$ had been localized inside the source $S2$ at $t=0$, it would have taken both routes and collapsed to being localized inside $D2$ at $t=8000$.

To analyze delayed-choice experiments, we need to define a model for causality and explain it's connection with causal intuition. Let us define the system as the two sources, the MZI, the two detectors, and the {neutron wavefunctions}. We can intervene on this system from outside the system. For example, we can intervene by sending a command to a source to emit a {wavefunction}. Let us define an intervention as a cause. Let us also define the effects as whatever is correlated with the intervention, {after adjusting for confounding variables using the \textit{do}-calculus}~\cite{Pearl,Woodward}. For example, the emission of a {wavefunction}, the motion of that {wavefunction} through the MZI, and the detection of that {wavefunction}. Many of our causal intuitions are based on interventions, not on temporal order. These causal intuitions are correlated with but not caused by temporal order. 

There are four possible ensembles of completed MZI experiments: (1) a neutron is emitted from $S1$ and detected in $D1$; (2) a neutron is emitted from $S2$ and detected in $D2$; (3) a neutron is emitted from $S1$ and detected in $D2$; and (4) a neutron is emitted from $S2$ and detected in $D1$. Repeated MZI experiments show that only ensembles 1 and 2 occur. Wheeler said this is "evidence that each arriving light quantum has arrived by both routes''~\cite{WheelerB}. Wheeler's thought experiments were done with photons, but the arguments are the same with neutrons.

Now consider a modified experiment where $B2$ is removed for the entire experiment. Repeated experiments show that ensembles 1, 2, 3, and 4 occur. Wheeler said either "one counter goes off, or the other. Thus the photon has traveled only \textit{one} route"~\cite{WheelerB}. 

Finally, consider a CF delayed-choice experiment where $B2$ is removed before each neutron is emitted at time $t=0$. At $t=5000$, we randomly choose to either reinsert or not reinsert $B2$. For the runs where we chose to not reinsert $B2$, we know that ensembles 1, 2, 3, and 4 occur. For the runs where we intervene to reinsert $B2$, only ensembles 1 and 2 occur. Wheeler said "Thus one decides the photon 'shall have come by one route, or by both routes' after it has \textit{already done} its travel'"~\cite{WheelerB}. How could a {quantum} at $B1$ know if an intervention will or will not occur before it reaches the point of intervention at $B2$? Wheeler said "we have a strange inversion of the normal order of time. We, now, by moving the beam-splitter in or out have an unavoidable effect on what we have a right to say about the \textit{already} past history of that photon"~\cite{WheelerB}. This is the presumed delayed-choice paradox.

Wheeler intended this as a demonstration that an interpretation in terms of a classical particle picture must use strange conceptual resources, like retrocausality. What happens in an interpretation in terms of a quantum wavefunction picture is the following: At $t=0$, we intervene on the system by sending a command to $S1$ or $S2$ to emit a neutron wavefunction. The wavefunction reaches $B1$ at $t=2000$, where half of it is transmitted towards $M1$ and the other half is reflected towards $M2$. These two halves reflect from $M1$ and $M2$, and then travel towards $D1$ and $D2$. If we do not intervene at $t=5000$, one half reaches $D1$ while the other half reaches $D2$ at $t=8000$, then one half collapses to a full wavefunction while the other half collapses to no wavefunction. If we do intervene at $t=5000$ by inserting $B2$, the two halves recombine at $B2$ and interfere constructively towards $D1$ and destructively towards $D2$ if the wavefunction came from $S1$, or vice versa if the wavefunction came from $S2$. Our intervention at $t=5000$ is uncorrelated with anything the wavefunction did for $0\le t<5000$. This means there is no delayed-choice paradox in the {quantum wavefunction picture of the CF delayed-choice experiment.} This has been explained before by Ellerman~\cite{Ellerman}.
\section{The Time-Reversed Formulation of the Delayed-Choice Experiment}\label{sec3}
Penrose pointed out that many quantum experiments can be explained equally well by a Time-Reversed Formulation (TRF) of quantum mechanics~\cite{PenroseA}. The TRF postulates that a single free particle with mass $m$ is described by a wavefunction $\phi^\ast(\vec{r},t)$ which satisfies the final conditions and evolves in time according to the time-reversed Schr\"{o}dinger equation:

\begin{equation}
-i\hbar\frac{\partial\phi^\ast}{\partial t}=-\frac{\hbar^2}{2m}\nabla^2\phi^\ast.
\label{1}
\end{equation}

We will use units where $\hbar=1$ and assume $\phi^\ast(\vec{r},t)$ is a traveling gaussian with a final standard deviation $\sigma=50$, momentum $k_x=0.4$, and mass $m=1$. 
\begin{figure}[htbp]
\centering
\includegraphics[width=5.0in]{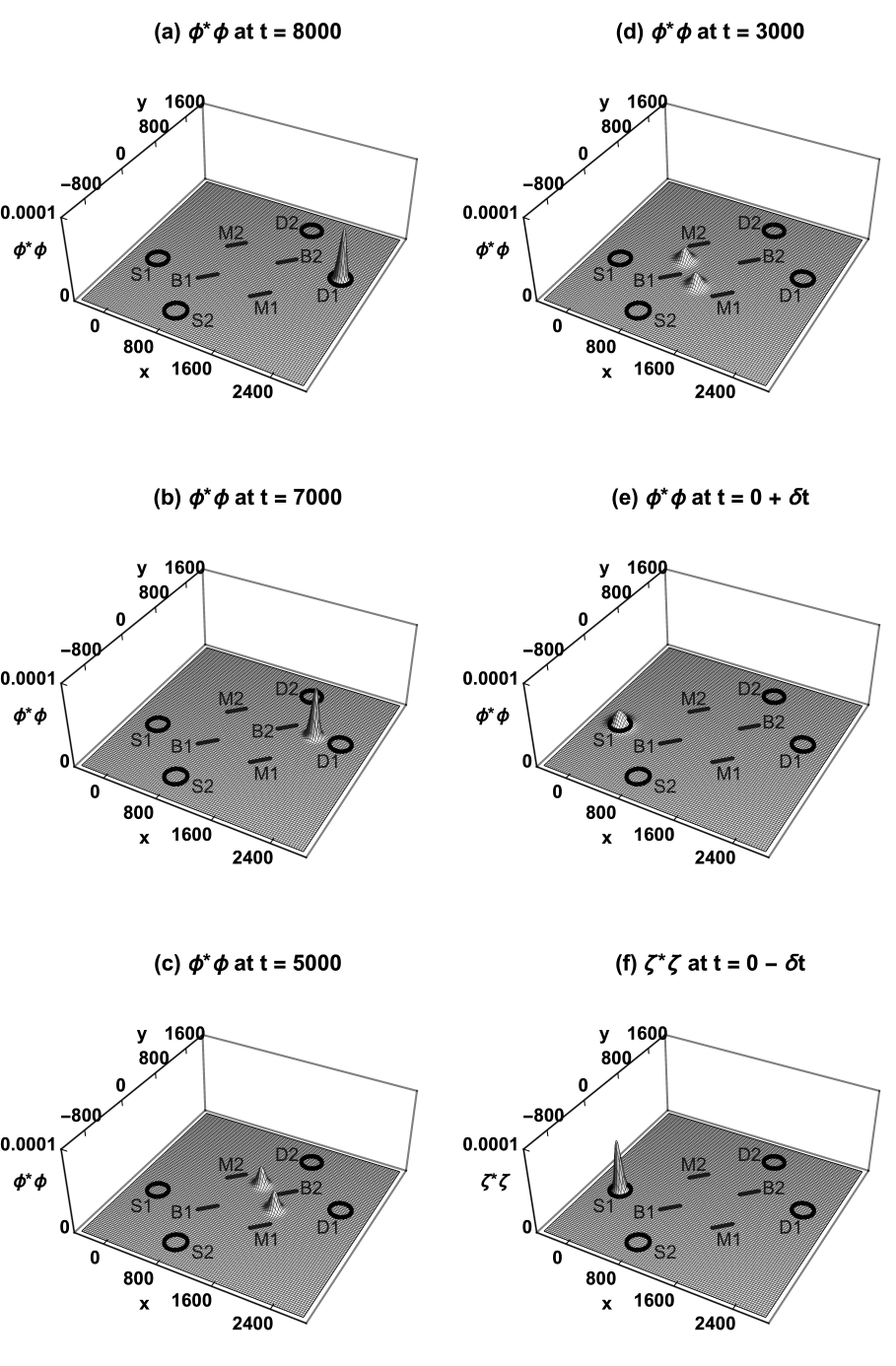}\label{fig3}
\caption{{The Time-Reversed Formulation} (TRF) of the MZI experiment, with a single neutron detected at $D1$. (\textbf{a}) The probability density $\phi^\ast\phi$ is localized inside $D1$. (\textbf{b}) $\phi^\ast\phi$ travels towards $B2$. (\textbf{c}) $\phi^\ast\phi$ has been split in half by $B2$. (\textbf{d}) The two halves have been reflected by $M1$ and $M2$. (\textbf{e})~The recombined $\phi^\ast\phi$ interferes constructively towards $S1$ and destructively towards $S2$, and $\phi^\ast\phi$ arrives at $S1$ but is not localized inside $S1$. (\textbf{f}) Upon preparation at $t=0$, $\phi^\ast$ collapses to a different wavefunction $\zeta^\ast$, localized inside $S1$. {Wavefunction collapse is a postulate of the TRF, and introduces a time-asymmetry.}}
\end{figure}
\mbox{Figure~\ref{fig3}} shows how the neutron's TRF probability density $\phi^\ast\phi$ evolves over time in the MZI, assuming the final condition is localization in detector $D1$. At time $t=8000$, we will assume $\phi^\ast\phi$ is localized inside the detector $D1$. At $t=7000$, $\phi^\ast\phi$ is traveling towards beam-splitter $B2$. At $t=5000$, $\phi^\ast\phi$ has been split in half by $B2$, and the halves are traveling towards mirrors $M1$ and $M2$. At $t=3000$, the two halves have been reflected by $M1$ and $M2$ and are both traveling towards $B1$. At $t=0+\delta t$, the two halves have been recombined by $B1$, with $\phi^\ast\phi$ interfering constructively towards source $S1$ and destructively towards source $S2$. Upon preparation at $t=0$, $\phi^\ast$ collapses to the different wavefunction $\zeta^\ast$, with $\zeta^\ast\zeta$ localized inside $S1$. Similarly, if $\phi^\ast\phi$ had been localized inside the detector $D2$ at $t=8000$, it would have taken both routes and collapsed to being localized inside $S2$ at $t=0$. {This cannot explain the MZI experiment: it implies the source of the neutron wavefunction will randomly be either $S1$ or $S2$, whereas we have full control over which source will emit the neutron wavefunction. This is an experiment that cannot be explained equally well by a TRF of quantum mechanics.}
\section{The Time-Symmetric Formulation of the Delayed-Choice Experiment}\label{sec4}
It is possible to explain delayed-choice experiments using a Time-Symmetric Formulation (TSF) of quantum mechanics. Time-symmetric explanations of quantum behavior predate the discovery of the Schr\"{o}dinger equation~\cite{Tetrode}, and TSF's have been developed many times over the past century~\cite{Friederich}. The TSF in this paper is a type IIB model, in the classification system of Wharton and Argaman~\cite{Wharton}. 

{In the CF and TRF, the beable is implicitly assumed to be a wavefunction which lives in configuration space and evolves in time. This is analogous to a point particle in Newtonian mechanics which lives in Euclidean space and evolves in time. In the TSF, the beable is explicitly assumed to be a transition amplitude density which lives in configuration spacetime. This is analogous to a world line in relativistic classical mechanics which lives in spacetime. Given an initial wavefunction $\psi(x,t)$ and a final wavefunction $\phi(x,t)$, the quantum amplitude $A$ to make the transition from $\psi(x,t)$ to $\phi(x,t)$ is \mbox{$A=\int_{-\infty}^{\infty}\phi^\ast(x,t)\psi(x,t)$dx}. The quantity $\phi^\ast(x,t)\psi(x,t)$ is the transition amplitude density. The transition probability $P$ is $P=A^\ast A$.}

Dirac showed that all the experimental predictions of the CF of quantum mechanics can be formulated in terms of transition probabilities~\cite{Dirac}. The TSF inverts this fact by postulating that quantum mechanics is a theory which experimentally predicts {only} transition probabilities. This implies the TSF has the same predictive power as the CF.

The TSF used in this paper has been described in detail and compared to other TSF's before~\cite{HeaneyA,HeaneyB}. Note in particular that the TSF used in this paper is significantly different than the Two-State Vector Formalism (TSVF)~\cite{TSVF}. First, the TSVF postulates that a {particle} is completely described by a two-state vector, written as $\langle\phi\vert\thickspace\vert\psi\rangle$. This two-state vector is not mathematically defined. In contrast, the TSF postulates that the {transition} of a particle is completely described by a complex transition amplitude density $\phi^\ast\psi$, which is mathematically defined. The TSF defines this transition amplitude density as the algebraic product of the two wavefunctions, which is a dynamical function of position and time. Second, the TSVF postulates that wavefunctions collapse upon measurement~\cite{ACE1}, while the TSF postulates that wavefunctions never collapse.

The CF and TRF postulate that a particle is described by one boundary condition and one wavefunction, while the TSF postulates that the transition of a particle is described by two boundary conditions and the algebraic product of two wavefunctions: a wavefunction $\psi(\vec{r},t)$ that obeys the Schr\"{o}dinger equation and satisfies only the initial boundary condition; and a time-reversed wavefunction $\phi^\ast(\vec{r},t)$ that obeys the time-reversed Schr\"{o}dinger equation and satisfies only the final boundary condition. The CF and TRF postulate that the wavefunction collapses instantaneously, indeterministically, and irreversibly into a different wavefunction at one of the boundary conditions, while the TSF postulates that wavefunctions never collapse. Consequently, the CF and TRF have intrinsic arrows of time, while the TSF has no intrinsic arrow of time.

\mbox{Figure~\ref{fig4}} shows how the absolute value of the neutron's TSF product wavefunction $\vert\phi^\ast\psi\vert$ evolves over time in the MZI experiment, assuming the initial condition is localization in source $S1$ and the final condition is localization in detector $D1$. At time $t=0$, $\vert\phi^\ast\psi\vert$ is localized inside the source $S1$. At $t=3000$, $\vert\phi^\ast\psi\vert$ has been split in half by beam-splitter $B1$, and the halves are traveling towards mirrors $M1$ and $M2$. At $t=5000$, the two halves have been reflected by $M1$ and $M2$ and are both traveling towards $B2$. At $t=7000$, the two halves have been recombined by $B2$, and the whole wavefunction travels towards detector $D1$. At $t=8000-\delta t$, $\vert\phi^\ast\psi\vert$ arrives at $D1$ and is localized inside $D1$. A second measurement at $t=8000+\delta t$ gives the same $\vert\phi^\ast\psi\vert$: there is no wavefunction collapse. Similarly, if $S2$ emits a neutron, it will always go to $D2$ via both routes. Wheeler could say this is "evidence that each arriving [neutron] has arrived by both routes"~\cite{WheelerB}. 

Now consider a modified TSF experiment where $B2$ is removed for the entire experiment. \mbox{Figure~\ref{fig5}} shows how the neutron's TSF product wavefunction $\vert\phi^\ast\psi\vert$ evolves over time in this modified experiment, assuming the initial condition is localization in $S1$ and the final condition is localization in $D1$. At $t=0$, $\vert\phi^\ast\psi\vert$ is localized inside source $S1$. At $t=3000$, $\vert\phi^\ast\psi\vert$ has been completely reflected by $B1$ and is traveling towards $M2$. At $t=5000$, $\vert\phi^\ast\psi\vert$ has been reflected by $M2$ and is traveling towards $D1$. At $t=7000$, $\vert\phi^\ast\psi\vert$ is still traveling towards $D1$. At $t=8000$, $\vert\phi^\ast\psi\vert$ arrives at $D1$ and is localized inside $D1$. A second measurement at $t=8000+\delta t$ would give the same $\vert\phi^\ast\psi\vert$: there is no wavefunction collapse upon measurement. The TSF product wavefunction $\phi^\ast\psi$ takes only the upper route. If the final state is changed to $D2$, the product wavefunction would take only the lower route. For any combination of source and detector, the product wavefunction always takes either the upper route or the lower route, never both routes. {Note that in contrast to the CF explanation of the same experiment, each run has a definite outcome, without a collapse and before and after a measurement is made. This solves one part of the measurement problem.} For this formulation, Wheeler's analysis is true: either "one counter goes off, or the other. Thus the [neutron] has traveled only \textit{one} route"~\cite{WheelerB}. 

Finally, consider a TSF delayed-choice experiment. When $B2$ is not present at times $0\le t<5000$ and not reinserted at $t=5000$, we infer that each product wavefunction always takes either the upper route or the lower route. When $B2$ is not present at times $0\le t<5000$, but we intervene to reinsert $B2$ for $5000\le t\le8000$, then each product wavefunction always takes both routes. Our intervention to reinsert $B2$ at $t=5000$ causes the product wavefunction to change from taking either route to taking both routes for $2000\le t<6000$: some of the effects occur {before} the intervention occurs, violating our causal intuition that effects never happen before interventions. {Note that this violation takes place at the quantum level, and we do not see macroscopic experimental evidence of a violation of causality.} There is a true delayed-choice paradox in the TSF delayed-choice experiment. Wheeler could say "we have a strange inversion of the normal order of time. We, now, by moving the [beam-splitter] in or out have an unavoidable effect on what we have a right to say about the \textit{already} past history of that [neutron]"~\cite{WheelerB}.
\begin{figure}[htbp]
\centering
\includegraphics[width=5.3in]{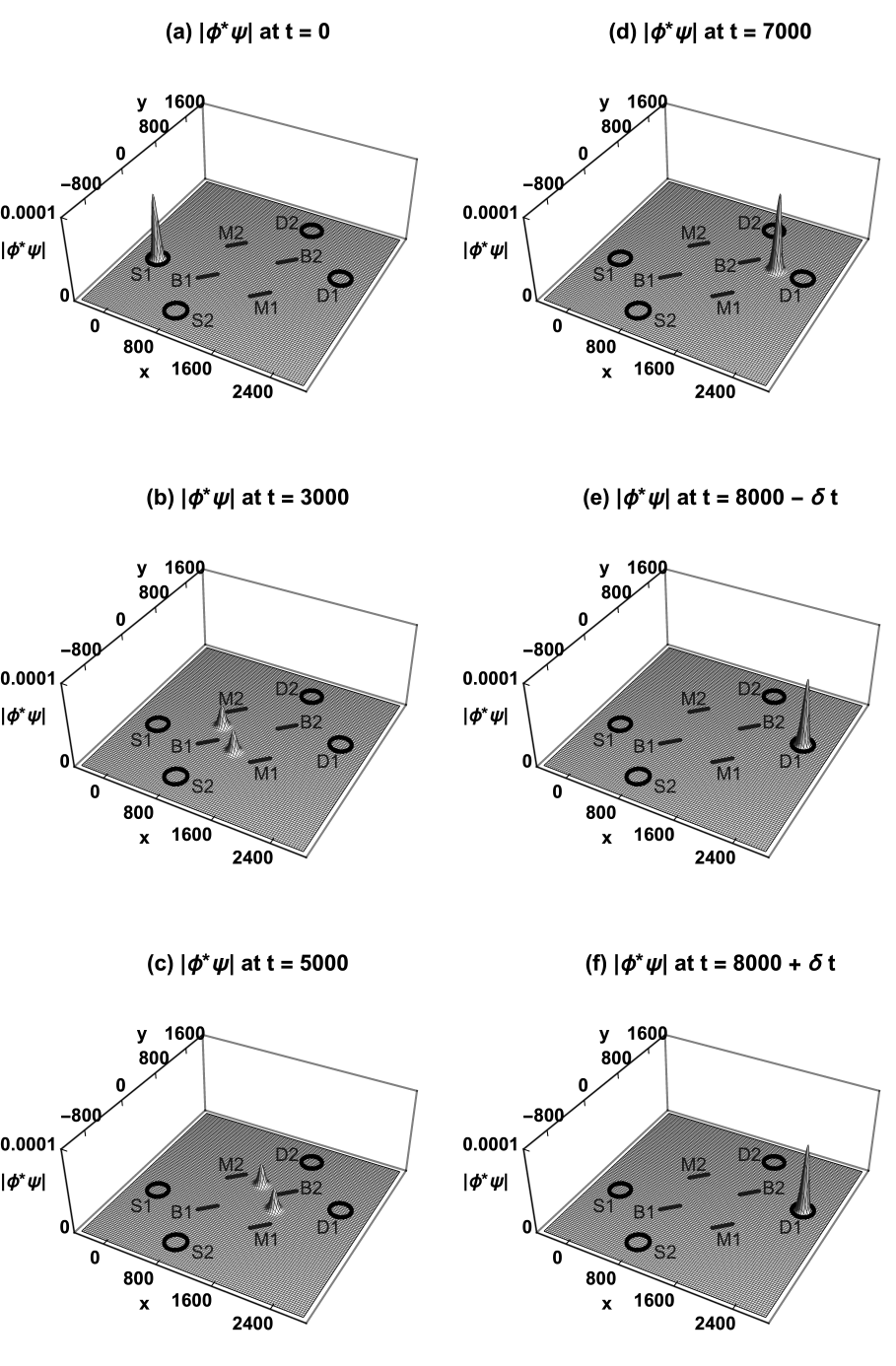}\label{fig4}
\caption{The Time-Symmetric Formulation of the MZI experiment. (\textbf{a}) The absolute value of the product wavefunction $\vert\phi^\ast\psi\vert$ is localized inside $S1$. (\textbf{b}) $\vert\phi^\ast\psi\vert$ is split in half by $B1$. (\textbf{c}) The two halves are reflected by $M1$ and $M2$. (\textbf{d}) The recombined $\vert\phi^\ast\psi\vert$ interferes constructively towards $D1$ and destructively towards $D2$. (\textbf{e}) $\vert\phi^\ast\psi\vert$ arrives at $D1$ and is localized inside $D1$. (\textbf{f}) A second measurement immediately afterward gives the same $\vert\phi^\ast\psi\vert$: there is no wavefunction collapse at any time. The probability for the transition is normalized to one.}
\label{fig4}
\end{figure}
\begin{figure}[htpb]
\centering
\includegraphics[width=5.0in]{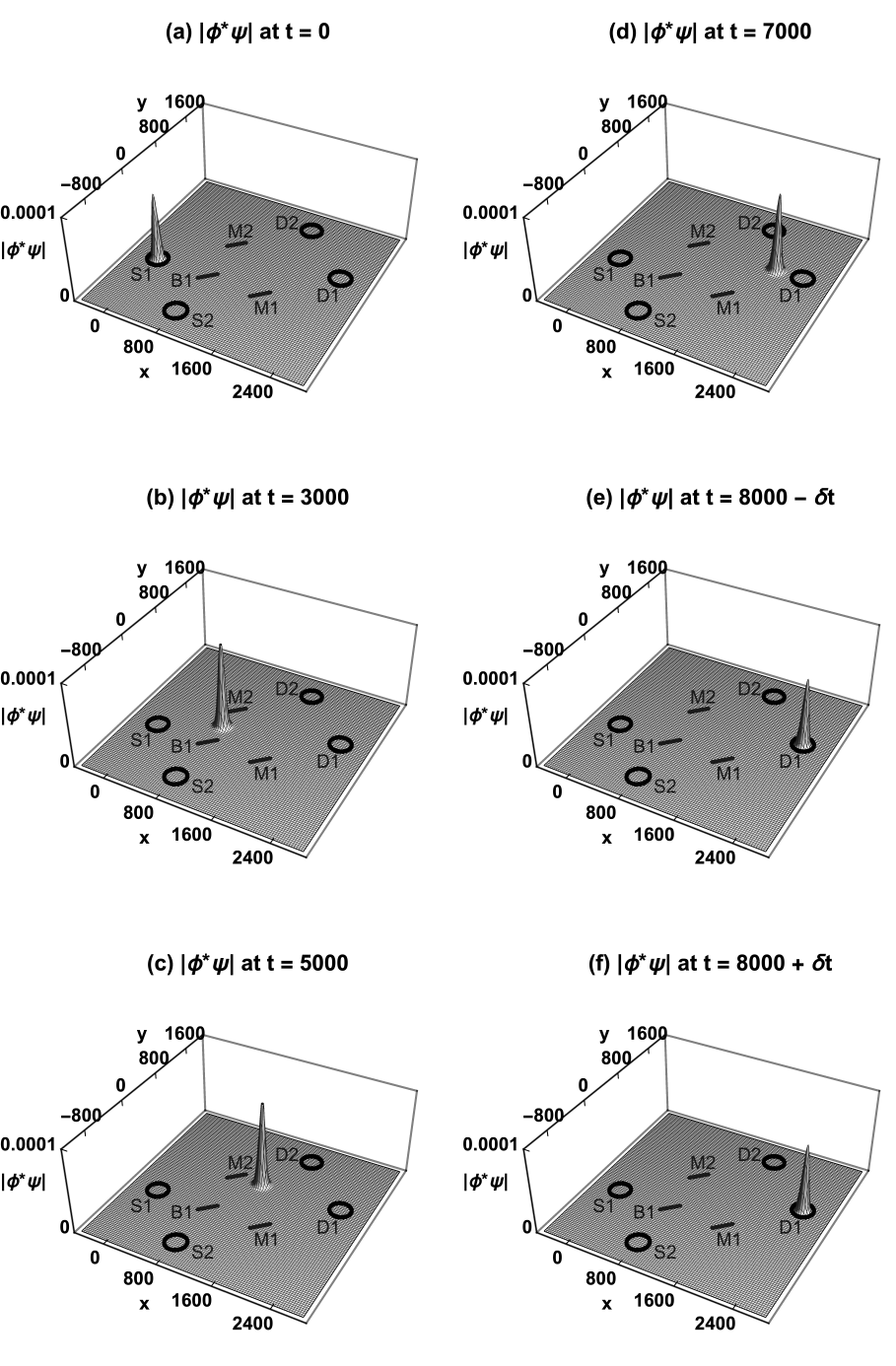}\label{fig5}
\caption{The Time-symmetric Formulation of a modified experiment with $B2$ absent. The particle is emitted by $S1$ and later detected by $D1$. (\textbf{a}) $\vert\phi^\ast\psi\vert$ is localized inside $S1$. (\textbf{b}) $\vert\phi^\ast\psi\vert$ is completely reflected by $B1$. (\textbf{c}) $\vert\phi^\ast\psi\vert$ is completely reflected by $M2$. (\textbf{d}) $\vert\phi^\ast\psi\vert$ travels towards $D1$. (\textbf{e}) $\vert\phi^\ast\psi\vert$ arrives at $D1$, and is localized inside $D1$. (\textbf{f}) A second measurement immediately afterward would give the same $\vert\phi^\ast\psi\vert$: there is no wavefunction collapse upon measurement. If the particle had been emitted by $S1$ and later detected by $D2$, $\vert\phi^\ast\psi\vert$ would have taken only the lower route.}
\end{figure}
\mbox{Figure~\ref{fig4}} shows how the absolute value of the particle's TF product wavefunction $\vert\phi^\ast\psi\vert$ evolves over time in the MZI experiment, assuming the initial condition is localization in source $S1$ and the final condition is localization in detector $D1$. At time $t=0$, $\vert\phi^\ast\psi\vert$ is localized inside the source $S1$. At $t=3000$, $\vert\phi^\ast\psi\vert$ has been split in half by beam-splitter $B1$, and the halves are traveling towards mirrors $M1$ and $M2$. At $t=5000$, the two halves have been reflected by $M1$ and $M2$ and are both traveling towards $B2$. At $t=7000$, the two halves have been recombined by $B2$, and the whole wavefunction travels towards detector $D1$. At $t=8000-\delta t$, $\vert\phi^\ast\psi\vert$ arrives at $D1$ and is localized inside $D1$. A second measurement at $t=8000+\delta t$ gives the same $\vert\phi^\ast\psi\vert$: there is no wavefunction collapse. Similarly, if $S2$ emits a particle, it will always go to $D2$ via both routes. Wheeler could correctly say this is \textquotedblleft evidence that each arriving [particle] has arrived by both routes" \cite{WheelerB}.  
\section{Discussion}\label{sec5}
The original Conventional Formulation (CF) explanations of the delayed-choice experiment by Lewis, Weizs\"{a}cker, and Wheeler only appeared to violate our causal intuition {at the quantum level.} The Time-Symmetric Formulation (TSF) explanation of the same experiment says the effects of an intervention can occur before the intervention, violating our causal intuition {at the quantum level.} Some may see this as reason to discard the TSF, but other aspects of quantum mechanics also violate our causal intuition (see papers in this issue). Perhaps there is something wrong with our causal intuition. What might be wrong?

The analyses in this paper suggest two overlapping considerations. First, our causal intuition that there is an arrow of time in the quantum world may be at fault~\cite{Price}. Humans live and develop intuitions in a macroscopic world with an omnipresent arrow of time set by the second law of thermodynamics and the low entropy past. It is natural that we would implicitly assume this arrow of time extends to the quantum world. However, the quantum world need not necessarily obey the second law. Consider the situation in \mbox{Figure~\ref{fig1}} where the neutron source $S1$ is a single excited nucleus capable of emitting a neutron, and the neutron detector $D1$ is a single ground state nucleus capable of absorbing a neutron. When the source nucleus decays, the neutron wavefunction travels through both arms of the MZI and is absorbed by the detector nucleus. The entropy during this transition does not change. There is then no thermodynamic arrow of time during the transition. {Our intuitive sense of an arrow of time can be attributed to the entropy gradient. However, we are usually not aware of this, and project time-asymmetry onto the temporal evolution of the wavefunction. This leads to unnecessary puzzles. As long as we do not have to speak of entropy there is no need to treat the two time directions as inequivalent, and this is a direct argument for the TSF. The fact that the TSF entails retrocausality should not concern us because that retrocausality occurs only at the quantum level and is compatible with our knowledge of the everyday world and scientific experiments.} Second, it is also ingrained in human intuition that nature is composed of objects which live in 3-dimensional space and evolve in time, because the velocities we experience are insignificant compared to the speed of light. However, the main lesson of the special theory of relativity is that nature is fundamentally ({3 + 1})-dimensional. Extending this lesson to the quantum world suggests analyzing experiments in a ({3N + 1})-dimensional configuration spacetime, where N is the number of degrees of freedom. {This is the quantum analog of the classical block universe viewpoint.} The TSF product wavefunction in ({3N + 1})-dimensional configuration spacetime then becomes the quantum equivalent of the world tube of a classical particle in ({3 + 1})-dimensional spacetime. This should allow a comparison of causation between the quantum and classical block universes. 

{The Conventional and Time-Reversed Formulations both require wavefunction collapse, which introduces a time-asymmetry. The Time-Symmetric Formulation has no wavefunction collapse, is time-symmetric, and gives the same experimental predictions for the delayed-choice experiment as the Conventional Formulation.} Heisenberg said "Since the symmetry properties always constitute the most essential features of a theory, it is difficult to see what would be gained by omitting them in the corresponding language~\cite{Heisenberg}". This is a significant advantage of the Time-Symmetric Formulation over the Conventional and Time-Reversed Formulations. In the Conventional and Time-Reversed Formulations the time-reversal operator is antilinear and double time reversal is not an identity transformation. In the Time-Symmetric Formulation the time-reversal operator is linear and double time reversal is an identity transformation~\cite{Watanabe}.

How might conventional causation be recovered in the classical limit? First, as the number of particles in a quantum system increases, the second law of thermodynamics comes into play, creating an effective arrow of time. Second, as the number of particles increases, the mean free distance between initial and final states decreases. This decreases the coherence length of the particle's product wavefunctions, so quantum phenomena which depend on delocalization are suppressed at macroscopic length scales. Third, as the quantum system interacts with the environment, decoherence effects will occur that make the quantum system behave in a more classical way~\cite{Zeh}.

Finally, the Time-Symmetric Formulation may have implications for cosmological boundary conditions. The wavefunction of the universe is believed to depend on the initial conditions at the Big Bang. If the Time-Symmetric Formulation is correct, the final conditions of the universe should play an equally important role in its evolution.

\vspace{6pt}

\end{document}